\documentclass[aps,prb,reprint,superscriptaddress]{revtex4-2}

\usepackage{graphicx}

\usepackage[version=4]{mhchem}
\usepackage{siunitx}
\usepackage{xcolor}
\usepackage[hidelinks]{hyperref}
\usepackage{relsize}
\usepackage{chemformula}
 
\draft 

\begin{document}


\title{Magnetic and geometrical control of spin textures in the itinerant kagome magnet \ce{Fe3Sn2}} 



\author{Markus Altthaler}\thanks{equal contribution}
\affiliation{Experimental Physics V, University of Augsburg, 86135 Augsburg, Germany}%
\affiliation{Department of Materials Science and Engineering, Norwegian University of Science and Technology (NTNU), 7043 Trondheim, Norway}
\affiliation{Center for Quantum Spintronics, Department of Physics, Norwegian University of Science and Technology (NTNU), 7491 Trondheim, Norway}
\author{Erik Lysne}\thanks{equal contribution}
\affiliation{Department of Materials Science and Engineering, Norwegian University of Science and Technology (NTNU), 7043 Trondheim, Norway}
\affiliation{Center for Quantum Spintronics, Department of Physics, Norwegian University of Science and Technology (NTNU), 7491 Trondheim, Norway}
\author{Erik Roede}
\affiliation{Department of Materials Science and Engineering, Norwegian University of Science and Technology (NTNU), 7043 Trondheim, Norway}
\author{Lilian Prodan}
\affiliation{Experimental Physics V, University of Augsburg, 86135 Augsburg, Germany}
\author{Vladimir Tsurkan}
\affiliation{Experimental Physics V, University of Augsburg, 86135 Augsburg, Germany}
\affiliation{Institute for Applied Physics, MD-2028, Chisinau, Moldova}
\author{Mohamed A. Kassem}
\affiliation{Department of Physics, Assiut University, 171516 Assiut, Egypt}
\author{Stephan Krohns}
\affiliation{Experimental Physics V, University of Augsburg, 86135 Augsburg, Germany}%
\author{Istv\'an K\'ezsm\'arki}
\email[]{istvan.kezsmarki@physik.uni-augsburg.de}
\affiliation{Experimental Physics V, University of Augsburg, 86135 Augsburg, Germany}%
\author{Dennis Meier}%
\email[]{dennis.meier@ntnu.no}
\affiliation{Department of Materials Science and Engineering, Norwegian University of Science and Technology (NTNU), 7043 Trondheim, Norway}
\affiliation{Center for Quantum Spintronics, Department of Physics, Norwegian University of Science and Technology (NTNU), 7491 Trondheim, Norway}



\date{\today}

\begin{abstract}
Magnetic materials with competing magnetocrystalline anisotropy and dipolar energies can develop a wide range of domain patterns, including classical stripe domains, domain branching, as well as topologically trivial and non-trivial (skyrmionic) bubbles. We image the magnetic domain pattern of \ce{Fe3Sn2} by magnetic force microscopy (MFM) and study its evolution due to geometric confinement, magnetic fields, and their combination. In \ce{Fe3Sn2} lamellae thinner than \SI{3}{\micro\meter}, we observe stripe domains whose size scales with the square root of the lamella thickness, exhibiting classical Kittel scaling. Magnetic fields turn these stripes into a highly disordered bubble lattice, where the bubble size also obeys Kittel scaling. Complementary micromagnetic simulations quantitatively capture the magnetic field and geometry dependence of the magnetic patterns, reveal strong reconstructions of the patterns between the surface and the core of the lamellae, and identify the observed bubbles as skyrmionic bubbles. Our results imply that geometrical confinement together with competing magnetic interactions can provide a path to fine-tune and stabilize different types of topologically trivial and non-trivial spin structures in centrosymmetric magnets.
\end{abstract}


\maketitle 

\section{INTRODUCTION} 
Progress in information and communication technology is driven by the discovery of new materials and phenomena that enable components with improved functionality. In a recent development, topologically protected spin textures, such as magnetic skyrmions and merons, are explored as functional nanoscale objects that can be utilized to process and store information \cite{Skyrme1962, Tomasello2014,Zhang2015}. Initially linked to the Dzyaloshinskii–Moriya interaction \cite{dzyaloshinskyThermodynamicTheoryWeak1958} (DMI), skyrmions were mostly studied in non-centrosymmetric crystals, such as cubic chiral \cite{muhlbauerSkyrmionLatticeChiral2009,munzerSkyrmionLatticeDoped2010,yuRoomtemperatureFormationSkyrmion2011} and axially symmetric polar magnets \cite{Kezsmarki2015, Butykai2017, Bordacs2017}. It was demonstrated that the emergence of skyrmions often is accompanied by peculiar electromagnetic properties like the topological Hall effect \cite{PhysRevLett.102.186602}, a zoo of collective excitations \cite{Garst_2017, Schwarze2015, ehlers2016}, pinning-free motion at ultra low current densities and polar dressing via the magnetoelectric effect \cite{Ruffe1500916}.

The formation of skyrmions and topologically equivalent skyrmionic bubbles \cite{yuMagneticStripesSkyrmions2011} is not restricted to non-centrosymmetric materials. In centrosymmetric magnets, such topologically non-trivial spin textures can be stabilized by frustrated exchange interactions \cite{okuboMultipleStatesSkyrmion2012}, originating from competing  dipolar interactions and magnetic anisotropy \cite{yuMagneticStripesSkyrmions2011,yuVariationTopologyMagnetic2017,wangCentrosymmetricHexagonalMagnet2016,tangTargetBubblesFe2020,montoyaTailoringMagneticEnergies2017,dingManipulatingSpinChirality2019,dingObservationMagneticSkyrmion2020}. In general, the spin texture of both skyrmions and skyrmionic bubbles is described by three parameters: the polarity ($p = \pm1$) corresponding to the direction of magnetiztion at the core, the helicity ($\gamma$) determined by the offset to the polar phase ($\Phi(\phi) = m\phi + \gamma$) and the vorticity
\begin{equation}
     m = \frac{1}{2\pi}\int_0^{2\pi}\text{d}\Phi(\phi),
\end{equation}
which counts the number of revolutions of the polar angle \cite{gobelSkyrmionsReviewPerspectives2020a}. Here, $\phi$ and $\Phi$ are the azimuthal angle in real space and the azimuthal angle of the local magnetization, respectively. In centrosymmetric magnets, bubbles with $m = 0$ are topologically trivial, whereas $m = 1$ indicates a topologically protected state, corresponding to skyrmions/skyrmionic bubbles. Depending on the relative magnitudes of the involved interactions, topologically trivial and non-trivial bubbles can even coexist, leading to complex mixed states with solitons of topological charge \cite{yuVariationTopologyMagnetic2017,loudonImagesBiskyrmionsShow2019}.

For achieving such exotic magnetic states and, hence, novel functional responses, materials with perpendicular magnetocrystalline anisotropy (PMA) and comparable values of uniaxial and shape anisotropy are particularly promising \cite{hubertMagneticDomainsAnalysis1998,yuVariationTopologyMagnetic2017}. The delicate balance of the two anisotropy terms creates substantial frustration, which enables the formation of unconventional magnetic order, including modulated magnetic structures, Bloch points and lines, as well as skyrmionic bubbles with opposite helicity \cite{Leonov2015}.

Here, we study \ce{Fe3Sn2}, a kagome layered-based itinerant magnet, where uniaxial ($K_u$) and shape anisotropies are comparable \cite{houManipulatingTopologyNanoscale2019}. The latter is described by the quality factor $Q = 2K_\text{u}/\mu_0M_\text{sat}^2$, which has a value of about 0.15. The material has a centrosymmetric rhombohedral structure formed of Fe-Sn bilayers alternating with Sn layers along the crystallographic \textit{c} axis. The anomalous Hall effect \cite{Kida2011,Wang2016, Ye2018}, the exceptionally high Curie temperature and massive Dirac fermions \cite{Ye2018} in the vicinity of the Fermi energy make \ce{Fe3Sn2} highly attractive for technological applications \cite{houObservationVariousSpontaneous2017,houCreationSingleChain2018}.
Geometric confinement has been demonstrated to have a strong impact on the nucleation of bubbles. More precisely, bubbles with $m = 0$ and $m = 1$ (type II and type I, respectively) are permitted to coexist or topologically trivial bubbles are completely expelled \cite{houManipulatingTopologyNanoscale2019}. Futhermore, by confining the geometry, Hou~\emph{et al.}~have demonstrated that the critical fields required to stabilize type I bubbles are significantly reduced, reaching realistic values for technological applications. Early studies on current-driven dynamics have also demonstrated that the helicity can be electrically tuned via spin transfer torque \cite{houCurrentInducedHelicity2020}.
The presented experiments by Lorenz transmission electron microscopy (LTEM), however, were limited to electron transparent samples with thicknesses $\mathsmaller{\lesssim}$ \SI{200}{\nano\metre}. In order to expand the research and systemically study the impact of the demagnetization energy for thicknesses varying over larger length scales, we apply magnetic force microscopy (MFM). We find that micrometer thick lamella-shaped samples exhibit bulk-like behavior, developing the peculiar dendrite pattern reported for \ce{Fe3Sn2} bulk samples \cite{heritageSpinReorientation2020,hubertMagneticDomainsAnalysis1998}. For smaller thickness ($\mathsmaller{\lesssim}$\SI{1}{\micro\meter}), we observe the formation of stripe and bubble domains which decrease in size with decreasing lamella thickness, following Kittel's law \cite{KittelScaling1946}. In addition, we study how variable magnetic fields transform the magnetic stripe domains into skyrmionic bubbles while keeping the sample thickness fixed and, vice versa, how the magnetic structure evolves as the sample thickness varies under constant magnetic field. The MFM measurements are complimented by micromagnetic simulation of the 3D magnetic order and its response to the applied manipulation, providing new insight into the physics of topological spin textures in magnetic materials with competing uniaxial and shape anisotropies.

\begin{figure}[htb]
\centering
\includegraphics[width=\linewidth]{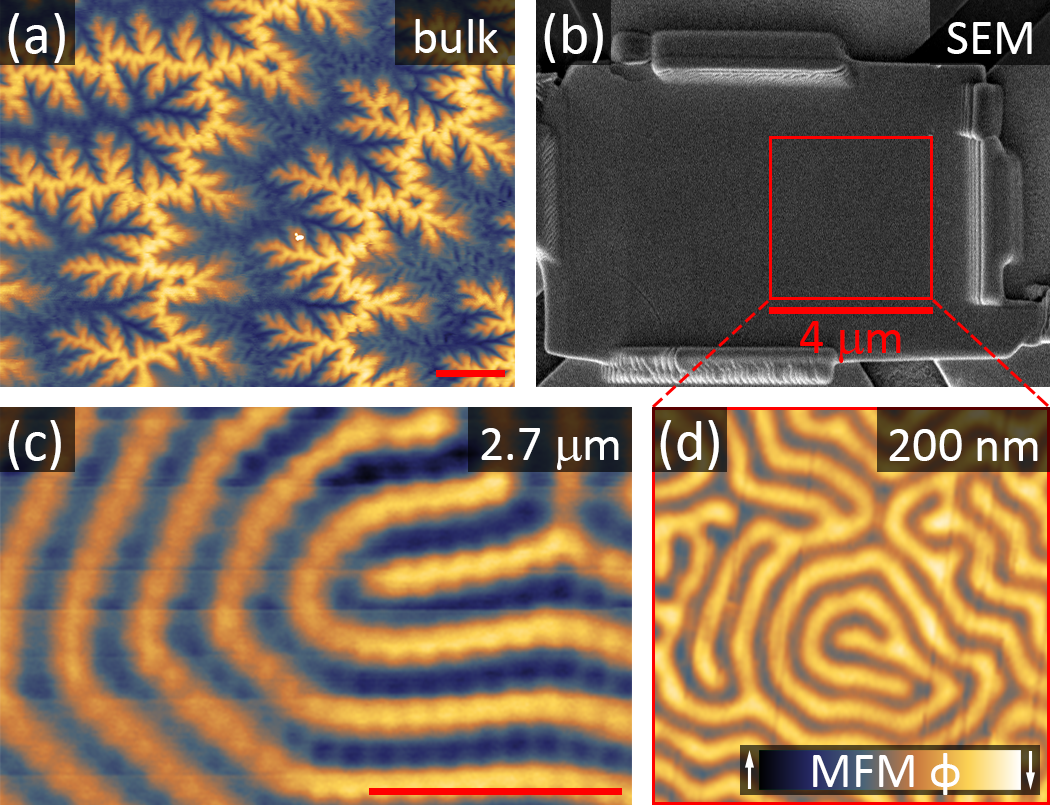}
\caption{(a) MFM imaging of a polished \textit{ab}-plane surface of a \ce{Fe3Sn2} single crystal reveals a peculiar dendrite pattern. (b) Representative SEM image showing a plan-view lamella cut from an \textit{ab}-plane surface as shown in (a).  (c) MFM image of a \SI{2.7}{\micro\meter} thick lamella revealing a pattern of out-of-plane magnetized alternating ferromagnetic stripe-like domains. Each individual stripe exhibits an undulated pattern characteristic for the onset of domain branching. (d) MFM scan of a \SI{200}{\nano\metre} thick lamella, showing a stripe-like domain pattern. All scale-bars (red) have a width of \SI{4}{\micro\metre}.  }
\label{fig1}
\end{figure}

\section{METHODS} 
\subsection{Crystal growth}
\ce{Fe3Sn2} single crystals are grown by the chemical transport reactions method. As starting material for growth, the preliminary synthesized polycrystalline powder is used. It has been prepared by solid state reactions from the high-purity elements: Fe (99.99\%) and Sn (99.995\%). Iodine is utilized as the transport agent in the single crystal growth. The growth is performed at temperatures between 730 and \SI{680}{\celsius}. Plate-like samples of thickness 20 - \SI{40}{\micro\meter} along the c-axis and 3-\SI{5}{\milli\meter} within the ab-plane are obtained.

\subsection{Lamella preparation}
\ce{Fe3Sn2} plan-view lamellae are prepared using a FIB-SEM at \SI{30}{\kilo\volt} ion beam acceleration voltage. We employ a Thermo Scientific G4UX DualBeam and a Zeiss Crossbeam 550 system equipped with an EasyLift EX NanoManipulator and a Kleindiek MM3A-EM micromanipulator, respectively. A triangular prism-shaped volume is isolated by milling undercuts at \SI{45}{\degree} to the surface  from a \ce{Fe3Sn2} single crystal. 
The specimen is then extracted using an micromanipulator and transferred to a copper half grid. By milling parallel to the original crystal surface, the samples are thinned to the desired thickness. The lamellae are then lifted off again and transferred to a substrate covered with \SI{200}{\nano\metre} of gold. In order to attach the lamella, the edges are welded to the surface using carbon or platinum as seen in the SEM image in Figure~1(b). Lamellae with controlled thicknesses between \SI{200}{\nano\meter} and several microns are prepared and fixed flat on a substrate. 

Following the same procedure, a prism-shaped volume is prepared for the wedge-like lamella. After polishing the original surface, the sample is transferred to a silicon substrate with the surface facing down. Mounted on a \SI{45}{\degree} stub, the prism is then cut from the apex at a low angle versus the base plane resulting in a wedge-like lamella with a thickness varying from \SI{400}{\nano\metre} to \SI{900}{\nano\metre} along the long axis.

\subsection{Magnetic imaging}
To investigate the magnetic domain textures \ce{Fe3Sn2}, we perform AC-MFM on an Oxford Instruments Cypher ES Environmental AFM at room temperature, mapping the phase shift of the read-out signal. For low-temperature and field-dependent studies, an attocube attoAFM I, equipped with a superconducting magnet is used, recording the frequency shift with a phase sensitive feedback loop. In both setups, the obtained image contrast is proportional to the projection of the gradient of the stray magnetic field to the tip magnetization \cite{KazakovaMFM2019}. In this work, we use Nanosensors TM PPP-MFMR probes magnetized prior to the scans.

The periodicity of the magnetic patterns in the MFM images is quantified using a MATLAB script. At each point of the MFM images the local periodicity is determined by fast Fourier transformation (FFT) over an area that is sufficiently large not to affect the obtained periodicity by a spatial cutoff. For wedged samples, the periodicity is averaged along lines of equal specimen thickness. 

\subsection{Micromagnetic simulations}
Micromagnetic simulations are performed using MuMax3 \cite{vansteenkisteDesignVerificationMuMax32014}. The energy terms included in the model represent the Heisenberg exchange, 1st order uniaxial anisotropy, Zeeman, and demagnetization energy terms,

\begin{equation}
\begin{split}
        \varepsilon = \int_{V_\text{s}} & \big[ A_\text{ex}(\nabla \mathbf{m})^2 - K_\text{u}m_z^2 \\
        & + M_\text{sat}\mathbf{B}_\text{ext}\cdot\mathbf{m} - \frac{1}{2}M_\text{sat}\mathbf{B}_\text{dem}\cdot\mathbf{m} \big] \text{d}\mathbf{r}.
\end{split}
\end{equation}
Here, $\mathbf{m}(x, y, z) \equiv \mathbf{M}(x, y, z)/M_\text{sat}$ is the reduced magnetization. The orientation of the magnetization $\mathbf{M}(x, y, z)$ within the sample volume $V_\text{s}$ is represented as a continuous unit vector field. $M_\text{sat}$ is the saturation magnetization, $A_\text{ex}$ is the exchange constant, $K_\text{u}$ is the 1st order uniaxial anisotropy constant, $\mathbf{B}_\text{ext}$ is the external magnetic field, and $\mathbf{B}_\text{dem}$ is the demagnetization field. Based on former studies at room temperature \cite{houObservationVariousSpontaneous2017,houCreationSingleChain2018,houManipulatingTopologyNanoscale2019,houCurrentInducedHelicity2020}, the magnetic easy axis is chosen to point along the \textit{z} axis. In order to avoid edge-induced effects, periodic boundary conditions are applied along the in-plane directions (\textit{x} and \textit{y}). Open boundary condition is used along the out-of-plane direction (\textit{z}), reflecting the constraints of \textit{ab}-plane cuts of varying thickness. The mesh is constructed by $\SI{8}{\nm}\times\SI{8}{\nm}\times\SI{8}{\nm}$ cells. The geometry is set to a square cuboid with length and width of $\SI{4 096}{\nano\meter}$. We investigate the thickness dependence of the magnetic patterns for a series of cuboids with different thicknesses. The magnetization is relaxed from an initial uniform out-of-plane magnetized state, where nucleation sites are included in the form of cylinders magnetized in the opposite direction to the surrounding matrix. The latter allows for stabilizing the maze-like ground state observed experimentally against uniform saturation.


\section{RESULTS}
\begin{figure}[htb]
\centering
\includegraphics[width=\linewidth]{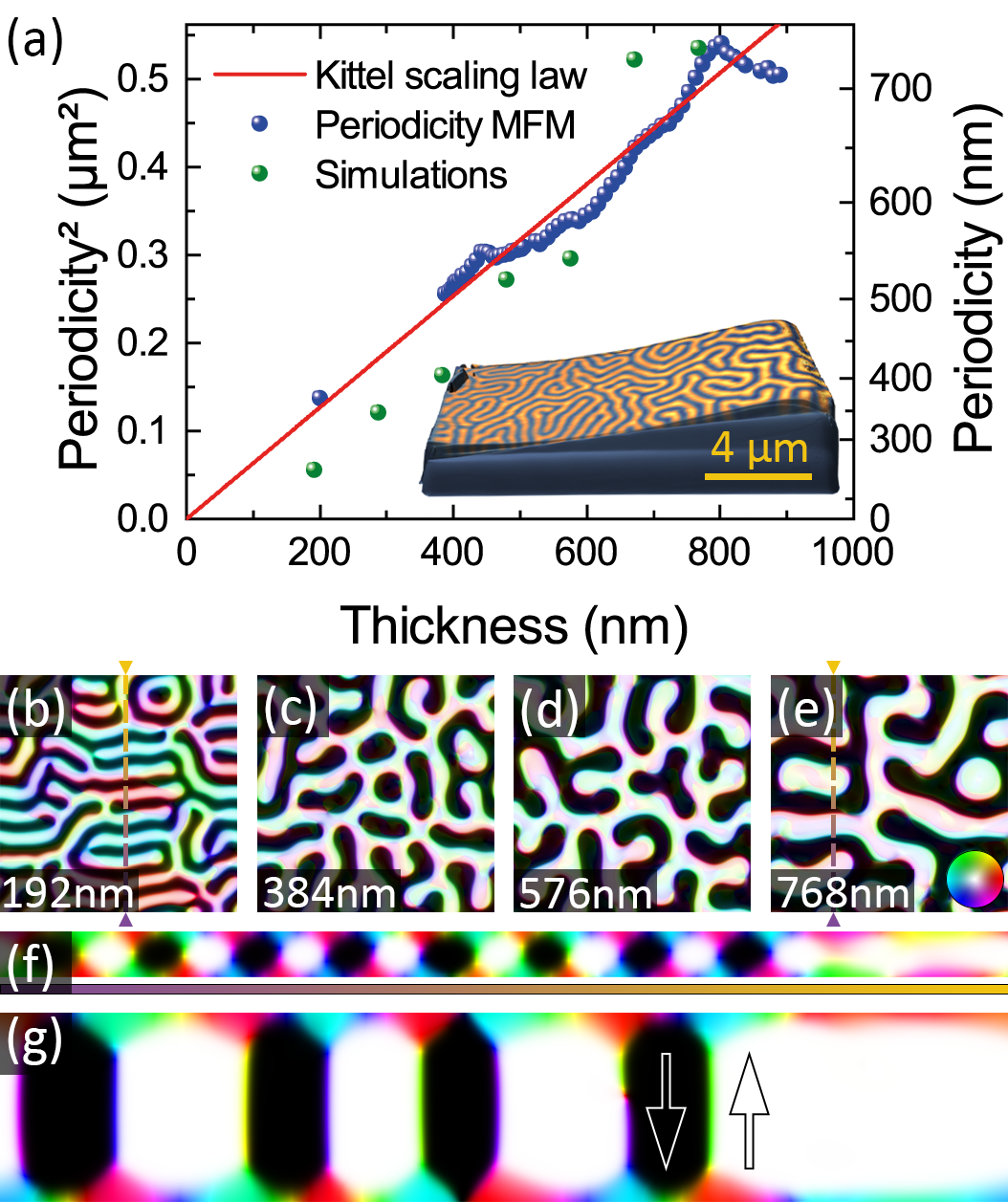}
\caption{(a) The inset displays the 3D visualization of the topography of the wedge-like lamella colored with the magnetic texture, where the bright and dark contrast corresponds to domains with magnetization up and down. The local periodicity (blue data points) is obtained via fast Fourier transformation (FFT) based analysis of the MFM signal and plotted versus the sample thickness. The analysis shows a square root dependence of the domain width, i.e., Kittel scaling (red line). A selection of simulated structures, representing a $4$x\SI{4}{\micro\metre\squared} area with periodic boundary conditions, is displayed in (b)-(e) where the sample thickness is \SI{192}{\nano\meter}, \SI{384}{\nano\meter}, \SI{576}{\nano\meter}, and \SI{768}{\nano\meter}. In the simulations brightness corresponds to the z-component of the magnetization, whereas colors encode the in-plane components in accordance with the color wheel in panel (e). The simulated trend in periodicity is in agreement with the experimental observations shown in (a), where the green data points show the periodicity derived from simulations scaled by a factor of 1.12 to match the scaling law of the experimental data. In (f) and (g) representative cross sections for the thin and thick end of the wedge lamella taken along the lines marked in (b) and (e) are shown. The color code is the same as in  with the simulations presented in top view.}
\label{fig2}
\end{figure}

MFM scans of the polished \textit{ab}-plane surface of a \ce{Fe3Sn2} crystal reveal a peculiar pattern of alternating bright and dark phase contrast, showing branched ferromagnetic domains, Fig.~1(a). Such domain branching is characteristic for \ce{Fe3Sn2} \cite{heritageSpinReorientation2020}, and materials with PMA in general \cite{hubertMagneticDomainsAnalysis1998}. The order of branching is expected to scale with the width of the underlying domains, vanishing with decreasing sample thickness \cite{hubertMagneticDomainsAnalysis1998}. Experimentally, we observe that the domain branching in \ce{Fe3Sn2} disappears at a sample thickness of about \SI{2.7}{\micro\meter}. The corresponding MFM image, Fig.~1(c), reveals predominantly stripe-like domains with undulated domain walls characteristic for the onset of domain branching. In contrast, a maze-like pattern of stripe-like domains with straight domain walls is observed in thinner lamellae as presented in Fig.~1(d), consistent with literature \cite{houObservationVariousSpontaneous2017}. The MFM data confirms that the applied FIB-based extraction procedure does not suppress or hinder the magnetic domain formation, confirming that the applied approach is suitable for preparation of high-quality \ce{Fe3Sn2} lamellae.


Figure~2 presents the effect of geometrical confinement. The possibility to utilize FIB milling to shape lamellae enables many opportunities to alter the geometry systematically and with nanometer precision. We seize this option to prepare a wedged lamella with thickness ranging from \SI{400}{\nano\metre} to \SI{900}{\nano\metre}, to gain insight into the continuous variation of the magnetic pattern with the sample thickness. 
The MFM image recorded on the wedged lamella is displayed as an inset to Fig.~2(a). Qualitatively, the alternating ferromagnetic stripe-like pattern resembles the magnetic domain state observed in the lamella in Fig.~1(d). Along the thickness gradient a clear widening of the stripes can be observed towards the thicker end of the lamella. Perpendicular to the gradient, no significant variation of the stripe width is found, pointing to a direct connection between the local thickness of the lamella and the width of the stripe domains. Our quantitative analysis shows a square root dependence of the domain width on the lamella thickness, i.e., the stripe width follows classical Kittel scaling \cite{KittelScaling1946}.

The variation of the stripe width versus sample thickness, revealed within the wedged lamella, provides a basis to refine the magnetic interaction parameters in \ce{Fe3Sn2} for more accurate micromagnetic simulations. Figures~2(b)-(e) show the results of our simulations for square cuboid lamellae with various thicknesses (\SI{192}{\nano\metre}, \SI{384}{\nano\metre}, \SI{576}{\nano\metre} and \SI{768}{\nano\metre}). With the parameter values $M_\text{sat} = \SI{566}{\kilo\ampere\per\meter}$, $A_\text{ex} = \SI{14.0}{\pico\joule\per\meter}$ and $K_\text{u} = \SI{30.0}{\kilo\joule\per\cubic\meter}$, the simulations quantitatively reproduce the thickness dependence of the stripe width (c.f. Fig.~2(a)), as observed in the surface region of the lamella. Importantly, the refined micromagnetic simulations provide an in-depth view on the 3D magnetization pattern. Figures~2(f) and (g) show cross-section images of the magnetization representative for the thin and thick end of the wedged lamella, respectively. In both cases alternating ferromagnetic domains, coaligned with the uniaxial anisotropy, constitute the core of the cuboids. In the surface region Néel capping occurs, explaining the gradual transition between domains in the MFM response. While the rotation of the magnetization is rather smooth in the surface region, the walls between up and down domains become considerably sharper in the internal region. This effect is strongest for the largest lamella thickness in Fig.~2(g).

\begin{figure}[b]
\centering
\includegraphics[width=\linewidth]{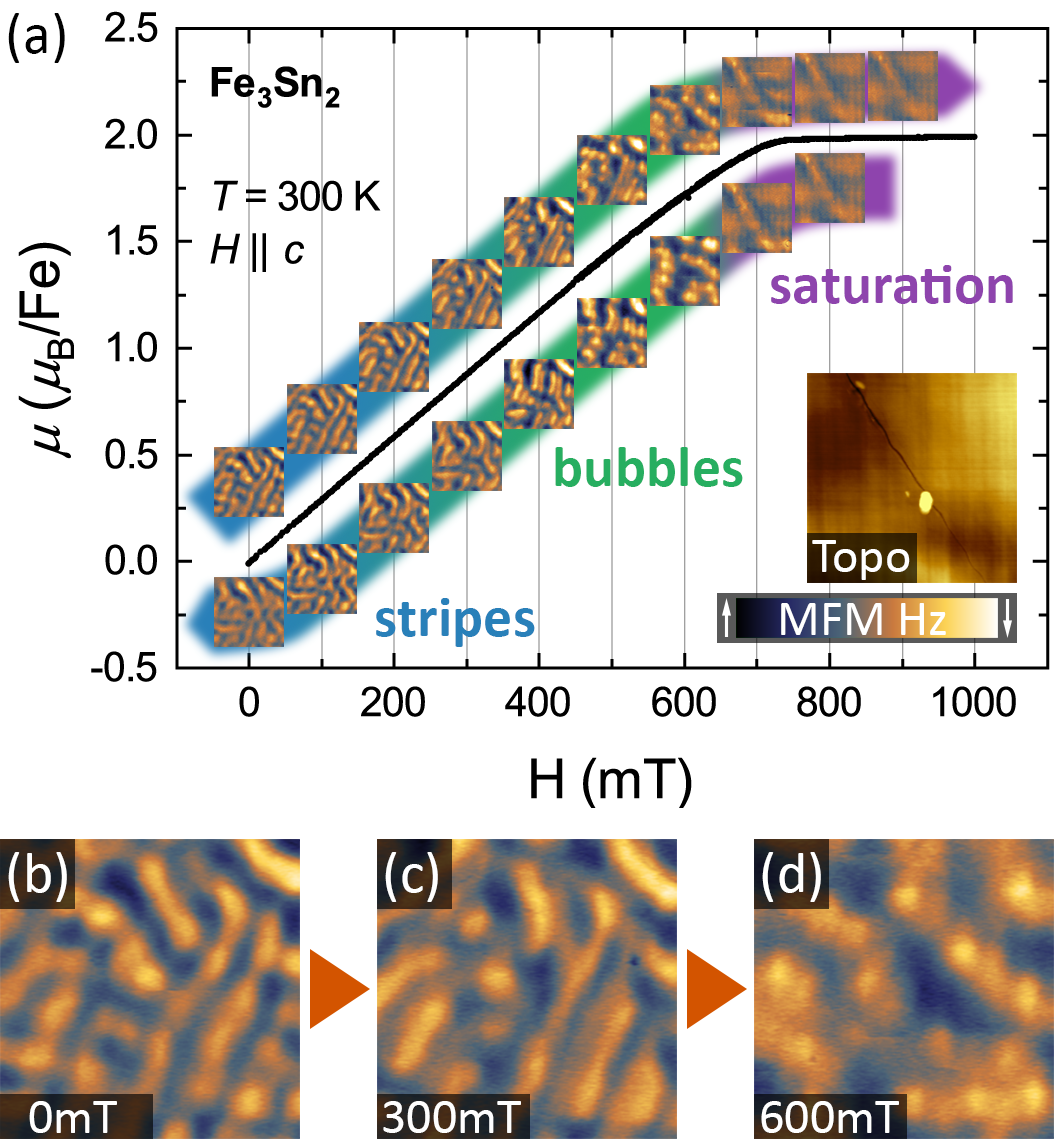}
\caption{(a) The magnetization data for a macroscopic sample (presented by the black data points) indicates that saturation magnetization is reached at about \SI{750}{\milli\tesla} and a hysteresis free cycle. Magnetic-field dependent MFM scans recorded on a \SI{450}{\nano\metre} thick lamella with stripe domains are presented as insets, revealing the local response to the applied field ($T$ = \SI{200}{\kelvin}). All images displayed show a $4$x\SI{4}{\micro\metre\squared} area with a \SI{25}{\hertz} color gradient. As the magnetic field increases, stripe domains (region marked by the underlying blue color) gradually transform to bubbles (green) before reaching the saturated state (purple). The residual contrast observed in the MFM scans obtained after \ce{Fe3Sn2} has been field polarized is attributed to the surface topography shown in the inset of (a). Upon decreasing the field, this evolution is reversed with spontaneously forming bubbles that subsequently elongate. The initially fragmented stripes grow and merge, relaxing back into a maze-like domain pattern. Panels (b)-(d) show three distinct field steps illustrating the field evolution in more detail.}
\label{fig3}
\end{figure}

After clarifying the thickness dependence of the magnetic texture, we investigate its magnetic field-driven evolution. Fig. 3(a) shows the field dependence of the magnetization for a macroscopic single crystalline sample at \SI{300}{\kelvin} with the magnetic field applied out of plane, i.e., parallel to the \textit{c} axis. In zero magnetic field, the sample has no remanent magnetization in agreement with the stripe-domain pattern observed by MFM and in the micromagnetic simulations. Following a linear increase, the magnetization saturates at ~\SI{750}{\milli\tesla} to the value of ~$2\,\mu_\textbf{B}$ per Fe atom. This value is in good agreement with earlier reports \cite{Caer1978, Malaman1978} and converts to $M_\text{sat}$= \SI{571}{\kilo\ampere\per\meter}, that agrees within 1\% with the value refined in the micromagnetic simulations. The detailed field evolution of the domain pattern is documented by a series of MFM images recorded on a \SI{450}{\nano\metre} thick \ce{Fe3Sn2} lamella at \SI{200}{\kelvin}. MFM images recorded over the same area in \SI{100}{\milli\tesla} steps are shown both for increasing the field up to \SI{900}{\milli\tesla} and decreasing it back to zero. With increasing magnetic field, the stripe domains gradually transform to bubble-like domains in between \SI{300}{\milli\tesla} and \SI{600}{\milli\tesla} before being expunged at the transition to the uniform ferromagnetic phase at approximately  \SI{700}{\milli\tesla}. Analogous results are observed at room temperature as discussed later. At high magnetic fields, a residual contrast is observed in the MFM scans independent of further variations in the field strength, which we attribute to topographic features shown in the inset to Fig.~3(a). With decreasing field, bubble-like domains form spontaneously just below \SI{700}{\milli\tesla} and subsequently elongate. Below \SI{500}{\milli\tesla} the fragmented stripes gradually grow and merge at low fields. 
The evolution of the magnetic pattern can be followed in more detail in Figs. 3(b)-(d), which  show three representative images recorded in zero field, \SI{300}{\milli\tesla} and \SI{600}{\milli\tesla}. In \SI{300}{\milli\tesla}, the area magnetized opposite to the field shrinks and some of the stripes break up into bubbles. In \SI{600}{\milli\tesla}, i.e., just below the saturation field, a pure bubble phase is realized. 
Thus, the behavior observed in the lamella is quantitatively similar to the domain evolution observed in classical ferromagnets with PMA, such as CoCr  and TbGdFeCo, under application of a magnetic field \cite{hubertMagneticDomainsAnalysis1998}. 
The emergence of completely different domain wall positions at zero field before and after ramping the magnetic field further reflects that no substantial pinning occurs in the \ce{Fe3Sn2} lamella. Thus, pronounced effects from defects, e.g., Ga implantation from the FIB preparation, can be excluded. \\
\begin{figure}[htb]
\centering
\includegraphics[width=\linewidth]{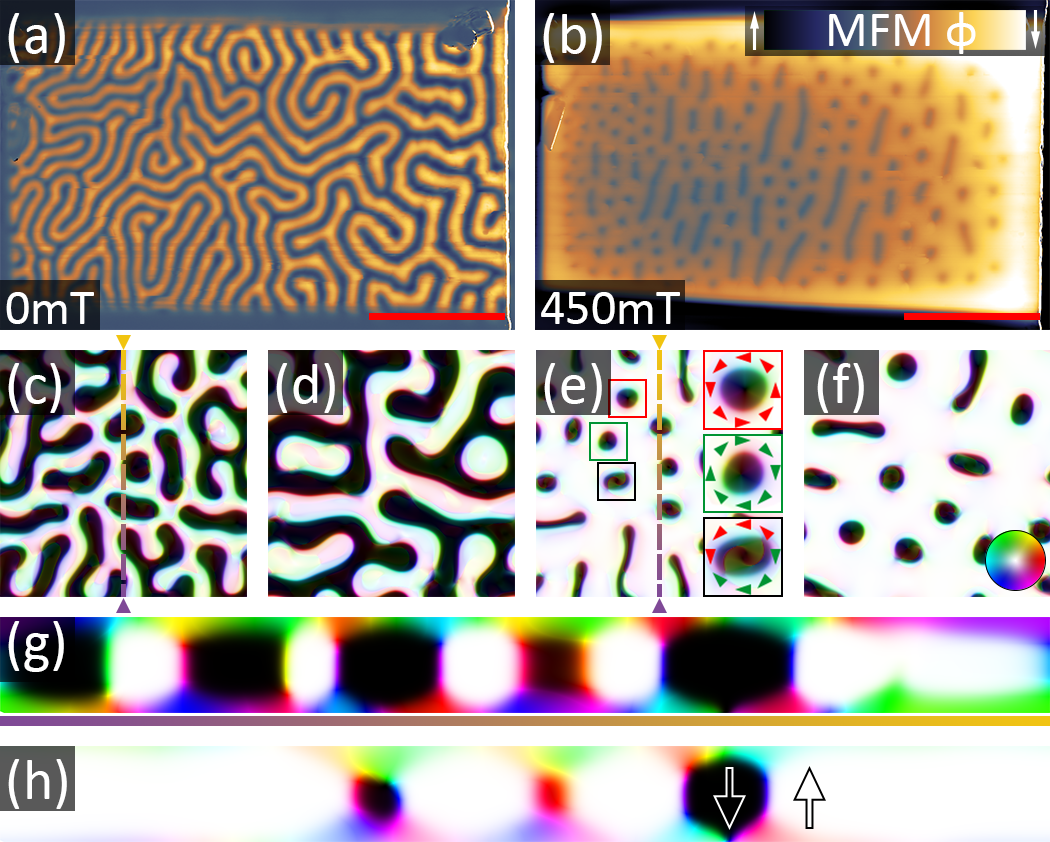}
\caption{ (a) The MFM image in zero field shows the gradual widening of the magnetic domains in a wedge-shaped lamella following Kittel's scaling law analogous to Fig. 2(a). (b) By applying a magnetic field of \SI{450}{\milli\tesla} along the out-of-plane direction, the micromagnetic texture can be driven toward the bubble domain phase. Both the remaining stripes as well as the induced bubbles follow Kittel scaling. (c), (d) Calculated magnetic texture in a region of 4x\SI{4}{\micro\metre\squared}, considering  a sample thickness of \SI{384}{\nano\meter} and \SI{768}{\nano\meter}, respectively, which represents the thin and thick end of the lamella in zero field. (e), (f) Calculated magnetic textures in external field of \SI{450}{\milli\tesla}, showing the coexistence of residual stripe domains and bubbles, reproducing the experimentally observed behavior. The insets to panel (e) show to coexistence of skyrmionic bubbles of opposite helicity (red, green), as well as trivial bubbles (black). (g), (h) show cross-sections illustrating the field evolution at the lines marked in (c) and (e). Both bubble domains and stripe domains are being clipped towards the surface by in-plane closure domains. All scale-bars (red) have a width of \SI{4}{\micro\metre}.}
\label{fig4}
\end{figure}
Next, we investigate the combined effect of thickness variation and magnetic field on the domain patterns in \ce{Fe3Sn2} by studying the domain structure in a wedged lamella as function of magnetic field. The experimental results and complementary micromagnetic simulations, displayed in Fig.~4, show the magnetic textures in zero field (a) and \SI{450}{\milli\tesla} (b). In zero field, stripes form a maze-like pattern, where the stripe width scales with the thickness according to Kittel's law, as presented in Fig.~2(a). In \SI{450}{\milli\tesla}, this pattern transforms to a state where shorter stripes coexist with bubbles. The nucleation of bubbles seems to start at the edges of the lamella, whereas the remaining stripes are concentrated within the internal part of the lamella. The phase coexistence of stripes and bubbles in \SI{450}{\milli\tesla} is consistent with the observations at \SI{200}{\kelvin}. Interestingly, we observe that not only the remaining fragmented stripe-like domains follow Kittel scaling, but also the diameter of the newly formed bubbles scales accordingly. 

Micromagnetic simulations representing the domain structure in the thin and the thick end of the lamella, respectively, are shown in Figure~4 (c) and (d). The simulations reproduce the domain pattern observed in zero field, including the thickness dependence of the stripe width. In 450 mT, similar to the experimental data, a mixed state of residual stripe-like domains and bubble domains is realized in the simulation, as shown in Figs. 4(e) and (f). The insets to panel (e) emphasize the coexistence of skyrmionic and topologically trivial bubbles, which cannot be concluded from the MFM scans alone. Furthermore, the simulated 3D magnetization reveals bubbles of opposite helicity.
A cross-section view of the thin end of the sample in zero field and \SI{450}{\milli\tesla}, Figs. 4 (g) and (h), show the magnetic texture in the core of the sample. In zero field, the pattern is similar to that found in the thick sample displayed in Fig.~2(g). In \SI{450}{\milli\tesla}, the area magnetized parallel to the field is extended and the remaining bubble domains of opposite magnetization are confined by sharp domain walls in the core and clipped at the surface by the in-plane closure domain pattern. As a consequence, the diameter of the bubble domains varies substantially between core and surface.

\section{SUMMARY}
In this work, we studied the response of the micromagnetic spin texture in the frustrated kagome magnet \ce{Fe3Sn2} under geometrical confinement and in magnetic field. Bulk samples reveal a branched \cite{hubertMagneticDomainsAnalysis1998} pattern of alternating ferromagnetic domains, which is characteristic for materials with PMA. In contrast, lamellae with thicknesses in the order of a few \SI{100}{\nano\meter} develop stripe patterns of alternating ferromagnetic up and down domains.  The thickness limit for the onset of domain branching is about \SI{2.7}{\micro\metre}. Below this limit, the size of the stripe domains follows classical Kittel scaling. The experimental results are corroborated by micromagnetic simulations, implying that the competition between dipolar interactions and uniaxial anisotropy is the driving force behind the formation and scaling of the domains. Furthermore, the micromagnetic simulations reveal Néel capping, i.e., the formation of closure domains at the surface. Analogous to the response to geometrical confinement, classical domain poling is also achieved by the application of external magnetic fields. We observe that stripe domains contract and break up into bubbles before a field-polarized single-domain state is reached. Intermediate $Q$-materials, like \ce{Fe3Sn2}, exhibit classical domain responses, thus enabling domain engineering following established strategies: introduction of geometrical confinement and application of external magnetic field. Importantly, this behavior also applies to both topologically trivial and non-trivial bubble domains. This yields the opportunity to tailor the host material to develop specific skyrmionic bubbles and mixed states, by optimizing domain size and densities for potential future applications. 

\section{Acknowledgments}
This work was supported by the DFG via the Transregional Research Collaboration TRR 80 From Electronic Correlations to Functionality (Augsburg/Munich/Stuttgart) and via the DFG Priority Program SPP2137, Skyrmionics, under Grant No. KE 2370/1-1 and by the project ANCD 20.80009.5007.19 (Moldova). E.L., E.R., and D.M. thank the Research Council of Norway for funding (project number 263228) and support through the Norwegian Micro- and Nano-Fabrication Facility, NorFab (project number 295864). M.A., E.L., and D.M. acknowledge support by the Research Council of Norway through its Centres of Excellence funding scheme, Project No. 262633, “QuSpin”. D.M. thanks NTNU for support via the Onsager Fellowship Program and the Outstanding Academic Fellows Program. 

%

\end{document}